\documentclass[reprint,
superscriptaddress,
 amsmath,amssymb,
 aps,
pra,
floatfix,
]{revtex4-2}

\usepackage{graphicx} % Required for inserting images
\usepackage{placeins}
\usepackage{multirow}
\usepackage{amsthm}
\usepackage{mathrsfs}
\usepackage[title]{appendix}
\usepackage[dvipsnames]{xcolor}
\usepackage{textcomp}
\usepackage{ulem}
\usepackage{booktabs}
\usepackage[caption=false]{subfig}
\usepackage[ruled,vlined]{algorithm2e}
\usepackage{listings}
\usepackage{enumitem}
\usepackage{cases}
\usepackage[useregional]{datetime2}
\usepackage[colorlinks=true, linkcolor=darkBlue, citecolor=darkBlue, urlcolor=darkBlue]{hyperref}
\usepackage{cleveref}
\usepackage{braket}
\usepackage{url}
\usepackage{dcolumn}% Align table columns on decimal point
\usepackage{bm}% bold math
\usepackage{comment}
\usepackage{orcidlink}

\definecolor{darkBlue}{RGB}{0,0,200}
\definecolor{darkGreen}{RGB}{13, 59, 2}
\allowdisplaybreaks

\newcommand{\llangle}{\langle\!\langle}
\newcommand{\rrangle}{\rangle\!\rangle}
\newcommand{\LLangle}{\left\langle\!\!\!\left\langle}
\newcommand{\NLLangle}{\Big\langle \!\!\!\!\! \Big\langle \,}
\newcommand{\RRangle}{\right\rangle\!\!\!\right\rangle}
\newcommand{\NRRangle}{\,\Big\rangle \!\!\!\!\! \Big\rangle}
\newcommand{\bw}{\boldsymbol{w}}
\newcommand{\bA}{\boldsymbol{A}}

\newcommand{\roruni}[2]{\href{https://ror.org/#2}{#1}}

\begin{document}
\preprint{APS/123-QED}

\title{Decoherence Cancellation through Noise Interference}

\author{Giuseppe D'Auria \!\orcidlink{0000-0001-7818-870X}}
\affiliation{International School for Advanced Studies (\roruni{SISSA}{004fze387}), Via Bonomea 265, 34136 Trieste, Italy}

\author{Giovanna Morigi \!\orcidlink{0000-0002-1946-3684}}
\affiliation{Theoretische Physik, \roruni{
Universit{\"a}t des Saarlandes}{01jdpyv68}, 
Campus E26, D-66123 Saarbr{\"u}cken, Germany}
\author{Fabio Anselmi \!\orcidlink{0000-0002-0264-4761}}
\affiliation{Department of Mathematics Informatics and Geoscience,\roruni{University of Trieste}{02n742c10}, Via Alfonso Valerio 2, Trieste, 34127, Italy}
\affiliation{\roruni{MIT}{042nb2s44}, 77 Massachusetts Ave, Cambridge, 02139, MA,USA.}
\author{Fabio Benatti \!\orcidlink{0000-0002-0712-2057}}
\affiliation{Department of Physics, \roruni{University of Trieste}{02n742c10}, Strada Costiera 11, 34014 Trieste, Italy}

\date{\today}
\begin{abstract}

\noindent
We propose a novel, feedback-free method to cancel the effects of decoherence in the dynamics of open quantum systems subject to dephasing. The protocol makes use of the coupling with an auxiliary system when they are both subject to the same noisy dynamics, in such a way that their interaction leads to cancellation of the noise on the system itself. 
This requires tuning the strength of the coupling between main and auxiliary systems as well as the ability to prepare the auxiliary system in a Fock state, which solely depends on the coupling strength. We investigate the protocol's efficiency to protect NOON states against dephasing in setups such as tweezers arrays of cold atoms. We show that the protocol's efficiency is robust against fluctuations of the optimal parameters and, remarkably, that it is independent of the temporal noise features. Therefore, it can be applied to cancel both Markovian and non-Markovian noise effects, in the regime where error-correction protocols become inefficient.
\end{abstract}

\maketitle

Loss of quantum coherence (decoherence) arises from the unavoidable interactions of quantum systems with their environments \cite{zurek2003decoherence} and is an outstanding challenge in the quest for next-generation quantum devices \cite{Preskill:2018}. A range of approaches has been proposed to improve the stability and reliability of quantum dynamics, thereby bringing the state-of-the-art closer to truly practical and scalable quantum technologies. These approaches include dynamical decoupling protocols, namely, open-loop quantum control techniques that are provably effective for Markovian processes~\cite{viola1999dynamical,khodjasteh2005fault,uhrig2008exact,liu2013noise}. Other approaches make use of weak and strong measurements, with success rates limited by the failure rates of the measurement processes involved~\cite{kondo2016using,bluhm2010enhancing}. 
Recently, protocols based on spectator qubits have been implemented that perform correction of phase errors during the circuit dynamics by measuring the states of qubits not involved in the quantum dynamics, but subject to correlated noise with computational qubits \cite{Singh:2023}. By combining measurements, data processing and feedforward operations, these correlated errors were suppressed during the execution of the quantum circuit. Similar ideas are at the basis of erasure conversion protocols in tweezers arrays of Rydberg atoms \cite{Scholl2023,Wu2022}.
The use of ancillary systems for robust quantum state preparation and computing is at the basis of a series of protocols, where entanglement between system and ancilla followed by measurements of the ancilla's state, or decay of the ancilla, can effectively pull the system into a fixed point of the composite dynamics \cite{Pielawa:2007,Giovannetti:2007,Morigi:2015,Menu:2022,Medina:2024,Alcalde:2024,cooper2024graph,periwal2021programmable}. An alternative ansatz is to prepare states which are immune from specific types of decoherence, ``dark states'' belonging to
decoherence-free subspaces, which are symmetry-protected from the external environment \cite{lidar1998decoherence,zanardi1997noiseless,bacon2000universal,kraus2008preparation}. This approach requires one to design appropriate coupling depending on the target state and on the environment \cite{kraus2008preparation,Reiter:2016}, and is typically effective for memoryless, Markovian reservoirs.

\begin{figure*}[t]
\centering
\begin{tikzpicture}
  \node[anchor=south west,inner sep=0] (image) at (0,0) {\includegraphics[width=0.95\linewidth]{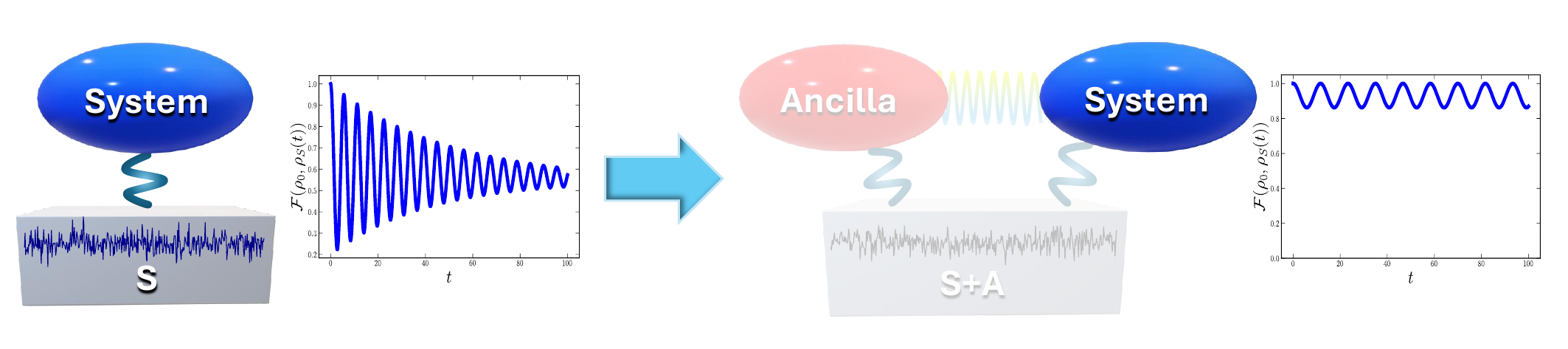}};
  \begin{scope}[x={(image.south east)},y={(image.north west)}]
    \node at (0.0,0.9) {\textbf{(a)}};
    \node at (0.45,0.9) {\textbf{(b)}};
  \end{scope}
\end{tikzpicture}
\caption{Schematic illustration of the proposed decoherence cancellation strategy. (a) The target system $S$ is initially subject to decoherence. (b) Upon enabling the interaction between $S$ and $A$ by bringing the ancilla into close proximity with the target system, both subsystems are exposed to a common correlated quantum noise.  Through appropriate preparation of the ancilla’s state and precise tuning of the interaction, a dark state is formed in the combined $S+A$ system, thereby canceling decoherence in $S$ and preserving only its unitary dynamics after tracing out the ancilla. }
   \label{fig:method}
\end{figure*}

In this Letter, we propose a protocol that achieves perfect noise cancellation of any quantum state of a system by the coupling with an auxiliary one, without requiring feedback operations, intermediate measurements, or symmetry constraints. 
The strategy is illustrated in Fig.\ \ref{fig:method}. It requires controlling the coupling between computational and spectator qubits that we refer to system $S$ and ancilla $A$.
By properly engineering the coupling between $S$ and $A$, the individual decoherence processes affecting each subsystem no longer simply add; instead, they can interfere destructively, yielding noise cancellation dynamics of the system.  Most importantly, we demonstrate that noise cancellation is achievable for any quantum state and temporal trajectory of the system and for any spectral property of the noise, as long as both the noise of system and ancilla is correlated.

To illustrate the procedure, consider a quantum system in a $d$-dimensional Hilbert space with Hamiltonian
\begin{equation}
\label{BoseHUbbardHamS}
H_S=\eta_S\, (J_S^z)^{2}\, -\, \gamma_S\,J_S^{x} - \Delta_S\, J_S^{z}\ ,
\end{equation}
where $\eta$, $\gamma_S$, $\Delta_S$ are real scalars and $J_S^{x,y,z}$ satisfy angular-momentum commutation relations, $\left[J_S^x\,,\,J_S^y\right]=\,i\,J_S^z$ and $(J_S^x)^2+(J_S^y)^2+(J_S^z)^2=\tfrac{d^2-1}{4}$. Let $\ket{\ell}_S$ be the eigenbasis of $J_S^z$, such that $J^z_S\ket{\ell}_S=\ell\ket{\ell}_S$, $-\frac{d-1}{2}\leq\ell\leq\frac{d-1}{2}$. Fluctuations of the parameter $\Delta_S$ give rise to decoherence effects which suppress off-diagonal terms with respect to the chosen basis. In the Markovian limit, these dephasing effects are described  by an irreversible, Markovian dynamics for the system density matrices $\rho_S(t)$ generated by a master equation of Gorini–Kossakowski–Sudarshan–Lindblad (GKSL) form~\cite{chruscinski2017brief},
\begin{align}
    \partial_t\rho_S(t)&=-i\Big[H_S\,,\,\rho_S(t)\Big]+\lambda\mathbb{M}[\hat J^z_S]\rho_S(t)\ ,
    \label{SMasterEq}
\end{align}
where $\mathbb{M}[\hat O]\hat\rho=\hat O\hat\rho\hat O^\dagger-\{\hat O^\dagger\hat O,\hat \rho\}/2$, and $\lambda$ is the decoherence rate. 
In this work, however, we consider a generic type of noise, which includes Markovian noise as a specific white-noise limit. For our discussion, we thus refer to a stochastic Hamiltonian
\begin{equation}
H_{S}'(t)=\,H_S\, +\, w_S(t)\,J^z_S\,,
\end{equation}
with $w_S(t)$ a generic centered stochastic potential, such that $\llangle w_S(t)\rrangle=0$, while higher moments are scalar functions determining the correlation functions of the noise. 

In order to mitigate decoherence, the system $S$ is coupled to an ancilla $A$. Here, for simplicity, we assume  that it is also a $d$ level quantum system with Hamiltonian
\begin{equation}
    \label{BoseHUbbardHamA}
H_A=\eta_A\, (J_A^{z})^2\, -\, \Delta_A\, J_A^{z} \,.
\end{equation}
In working experimental conditions, the ancilla is also affected by the same noise as the target system. Assume the $SA$ interaction has the form $H_{SA}^{I} = \alpha \,J^z_S\otimes J^z_A$, where $\alpha\in\mathbb{R}$ scales the strength.
Let us suppose that, by switching on the interaction, the noise affects the compound system $S+A$ in a correlated way as described by the stochastic Hamiltonian $H_{SA}(t)=H_0+H_{\rm stoch}(t)$, where $H_0\,:=\, H_S\otimes\mathbb{I}_A\,+\,\mathbb{I}_S\otimes H_A\,+H_{SA}^{I}$ is a deterministic contribution and the stochastic term is given by
\begin{align}
   \nonumber
    H_{\rm stoch}(t)&= w_S(t)\,J^z_S\otimes \mathbb{I}_A\,+\,w_A(t)\, \mathbb{I}_S\otimes J^z_A\\ 
     \label{HSA}
     &+\,w_{SA}(t)\,\alpha J^z_S\otimes J^z_A\, ,
\end{align}
where $\alpha$ is a time-independent parameter. Here, the first two are stochastic potentials affecting independently $S$ and $A$ whereas the last term describes a stochastic interaction between system and ancilla. Thus,  preparing the auxiliary system in an eigenstate of $J_A^z$, say, $J_A^z|\ell\rangle_A=\ell|\ell\rangle_A$, then for any state $|\Psi\rangle_{SA}=|\psi\rangle_S\otimes |\ell\rangle_A$ of the composite system the following relation holds
$$
H_{\rm stoch}(t)\vert\Psi\rangle_{SA}=\Big(w_S(t)+\alpha\ell w_{SA}(t))\Big)J^z_S\otimes\mathbb{I}_A\vert\Psi\rangle_{SA}
$$
plus a contribution of the form $\ell\,w_A(t)\vert\Psi\rangle_{SA}$.
Suppose that $w_{SA}(t)=c_{SA}\,w_S(t)$ with a real proportionality factor $c_{SA}\in \mathbb{R}$; then, remarkably, 
cancelling the noise affecting the system $S$ only requires to tune the interaction strength $\alpha$ such that $\alpha=-1/(\ell \,c_{SA})$. For this value, $H_{\rm stoch}(t)|\Psi\rangle_{SA}=\ell\,w_A(t)\vert\Psi\rangle_{SA}$ and the system $S$ is protected against the detrimental effects of the environment. 
Of course, noises are typically not controllable and thus hardly considerable to be proportional; instead, what can be controlled are their correlations.
This will be explicitly  elaborated on for the case of white-noises whereby, as shown in Sections~\ref{SM:app1} and~\ref{SM:ME} of the SM, averaging over the noise yields a GKSL master equation  for the density matrices $\rho$ of the system $S$.
Notice however that the above heuristic argument indicates the  ability of our proposal to cancel decoherence not only for colored noises with time-correlations manifesting memory effects, but, potentially also for $1/f$ noises, that present error-correction protocols cannot address (see Section~\ref{SM:Dech_canc} of SM).

Assume the correlations of the white-noises $w_S(t),w_A(t)$ and $w_{SA}(t)$ in the Hamiltonian~\eqref{HSA} satisfy $\llangle w_i(t)w_j(s)\rrangle=\Lambda_{ij}\delta(t-s)$ with correlation matrix
\begin{equation}
\label{corr-mat}
  \Lambda=[\Lambda_{ij}]=\lambda\begin{pmatrix}
  1&1&\alpha\\
  1&1&\alpha\\
  \alpha&\alpha&\alpha^2
  \end{pmatrix}\ ,
  \end{equation}
that is positive semi-definite for $\lambda>0$ and $\alpha \in {\mathbb{R}}$, see \cite{ghirardi1990relativistic, bassi2003dynamical} and Section~\ref{SM:app1} of the SM.
Averaging over the noises, these conditions yield a Lindblad master equation $\partial_t\rho_{SA}(t)=\mathbb{L}\rho_{SA}(t)$ for the density matrix of S and A. Its GKSL generator is of the form $\mathbb{L}=\sum_{i,j=1}^3\mathbb{L}_{ij}$, where 
\begin{equation}
\label{gen-split}
\mathbb{L}_{ij}\rho_{SA}=\Lambda_{ij}\Big(A_i\,\rho_{SA}\,A_j-\frac{1}{2}\Big\{A_j\,A_i\,,\,\rho_{SA}\Big\}\Big)\ .
\end{equation}
It proves convenient to recast it as
\begin{eqnarray}
\nonumber
\mathbb{L}\rho_{SA}(t)&=&-i\big[ H_0,\rho_{SA}(t)\big]+\sum_{j=1}^2\mathbb{M}[\hat O_j]\rho_{SA}(t)\\
 &&+\left(
        \mathbb{M}[\hat O_{+}]
        -\mathbb{M}[\hat O_{-}]
     \right)\rho_{SA}\ ,
\label{rodtgenSA2.0b}
\end{eqnarray}
where $\hat{O}_1=\sqrt{\lambda}J_S^z\otimes J_A(\alpha)$, $\hat{O}_2=\sqrt{\lambda}J_S(\alpha)\otimes J_A^{z}$, and $\hat O_{\pm}=\sqrt{\frac{\lambda}{2}}\Bigl(J_S^{z}\otimes\mathbb{I}_A\pm\mathbb{I}_S\otimes J_A^{z}\Bigr)$, such that ($j=S,A$)
\begin{equation}   J_j(\alpha)=\mathbb{I}_j+\alpha J^z_j\,.
\end{equation}
This form allows us to recovers Eq.~\eqref{SMasterEq} after setting $\alpha=0$ and tracing over the ancilla Hilbert space.

Now, assuming that the auxiliary system $A$ is in the state $P^\ell_A=|\ell\rangle\langle\ell|$, one can choose
$\displaystyle \alpha=-1/\ell$, so that  $J_A(\alpha)P^\ell_A=P_A^\ell J_A(\alpha)=0$. This condition leads to
\begin{eqnarray}
    \label{gens1}
-i[H_0,\rho_{SA}]&=&-i[H_S-J^z_S,\rho_S]\otimes P^\ell_A\ ,\\
\label{gens2}
\mathbb{M}[\hat O_1]\rho_{SA}&=&\mathbb{M}[\hat O_2]\rho_{SA}=0\\
\label{gens3}
\big(\mathbb{M}[\hat O_+] &-& \mathbb{M}[\hat O_-]\big)\rho_{SA}=0\, .
\end{eqnarray}
As a result, the superoperator on the right-hand side of~\eqref{rodtgenSA2.0b} acts simply as
\begin{equation}
\mathbb{L}\Big(\rho_S\otimes P^\ell_A\Big)=-i[H_0-J^z_S,\rho_S]\otimes P^\ell_A\ \,.
    \label{gens2.1}
\end{equation}
Consequently, the dynamics $\displaystyle\gamma_t={\rm e}^{t\mathbb{L}}$ generated by the exponential action of $\mathbb{L}$ is unitary 
and leaves the ancilla unaffected:
\begin{equation}
\label{coupledGKSL8}
\gamma_t\Big(\rho_S\otimes P^\ell_A\Big)={\rm e}^{-itH_S^{eff}}\,\rho_S\,{\rm e}^{itH_S^{eff}}\otimes\ P^\ell_A\ ,
\end{equation}
where $H_S^{eff}=H_S-J^z_S$.
Notice that, under the above conditions, the purely Hamiltonian dynamics in~\eqref{coupledGKSL8} holds true \textit{for all} system's initial states $\rho_S$.

\begin{figure}[t]
\centering
\includegraphics[width=\linewidth]{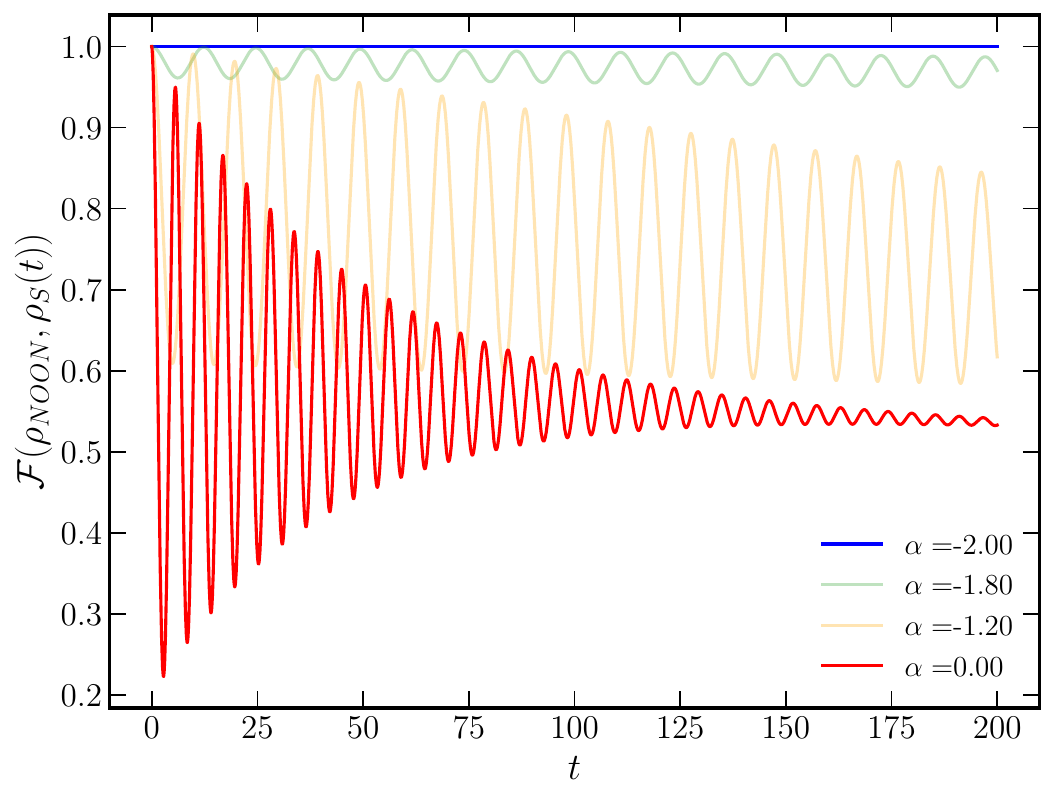}
\caption{Fidelity of staying NOON at time $t$ for a system $S$ with one qubit, $N_S=N_A=1$, with fixed $l=1/2$, for different values of $\alpha$. Optimal cancellation of decoherence is achieved for $\alpha = -1/l = -2$. Here we used $\gamma_S =0.5$, $\gamma_A=0$, and  $\Delta_S =\Delta_A=-1$, which also guaranties cancellation of the residual unitary oscillations.  }
   \label{fig:dech_qubits}
\end{figure}

Interactions between main and auxiliary system, such as the ones discussed here, could be implemented in platforms including main and spectator qubits, such as, for instance, in~\cite{Singh:2023}, as well as atomic tweezers for atoms inside a resonator~\cite{Hartung:2024}, for Rydberg arrays~\cite{srakaew2023subwavelength,scholl2023erasure,schlosser2020assembled,anand2024dual,bornet2023scalable}, including hybrid lattice-tweezer platforms~\cite{tao2024high}. In the tweezers setups, the correlated noise could be for instance a state-dependent fluctuation of the tweezers trap potential that the atoms experience. Crucial for the viability is the capability to prepare the auxiliary system in a Fock state and to control the system-ancilla interaction. While there is an optimal working point, in what follows we show that the protocol's efficiency is robust against fluctuations of the optimal parameters. For this purpose, we discuss the protocol efficiency for the stabilization of NOON states against dephasing noise. 
We consider two double-well traps for Bosons with long-range interactions~\cite{landig2016quantum,habibian2013bose,ritsch2013cold,klinder2015observation}: one double well representing the system $S$ and the other the ancilla $A$, where in the double well corresponding to the system $S$, a NOON state has been prepared, $\ket{\Psi_S^\pm}={\ket{N,0}_S\pm\ket{0,N}_S}/{\sqrt{2}}$, where $|N,0\rangle_S$ is the state of $N$ Bosons in the left well and $0$ in the right, while $|0,N\rangle$ is the state with the swapped occupations.  It satisfies $J^z_S\ket{\Psi_S^\pm}=-\frac{N}{2}\ket{\Psi^\mp_S}$ and $(J^z_S)^2\ket{\Psi_S^\pm}=\frac{N^2}{4}\ket{\Psi_S^\pm}$. In the presence of dephasing noise, the stochastic Hamiltonian of system and ancilla can be reduced to the form $H_{SA}'=H_{SA}+\frac{1}{2}\alpha\, w_{SA}(t)K_{SA}$ with $\alpha$ the coupling strength between the two double wells and $K_{SA}=(J_S^z)^2\otimes \mathbb{I}_A  + \mathbb{I}_S\otimes (J_A^z)^2$, see Section~\ref{App:case-study} of the SM. 
For $\alpha=0$, the stochastic Hamiltonian $H_{SA}'$ predicts that quantum coherence is lost over a time-scale $1/\lambda$. If instead the ancilla is prepared in the eigenstate $\ket{\ell}_A$ and $\alpha_c=-1/\ell$ , the effects of noise cancel exactly $\alpha=-1/\ell$, as visible in Figure~\ref{fig:dech_qubits}. 

\begin{figure*}[t]
\centering
\begin{tikzpicture}
  \node[anchor=south west,inner sep=0] (image) at (0,0) {\includegraphics[width=0.98\linewidth]{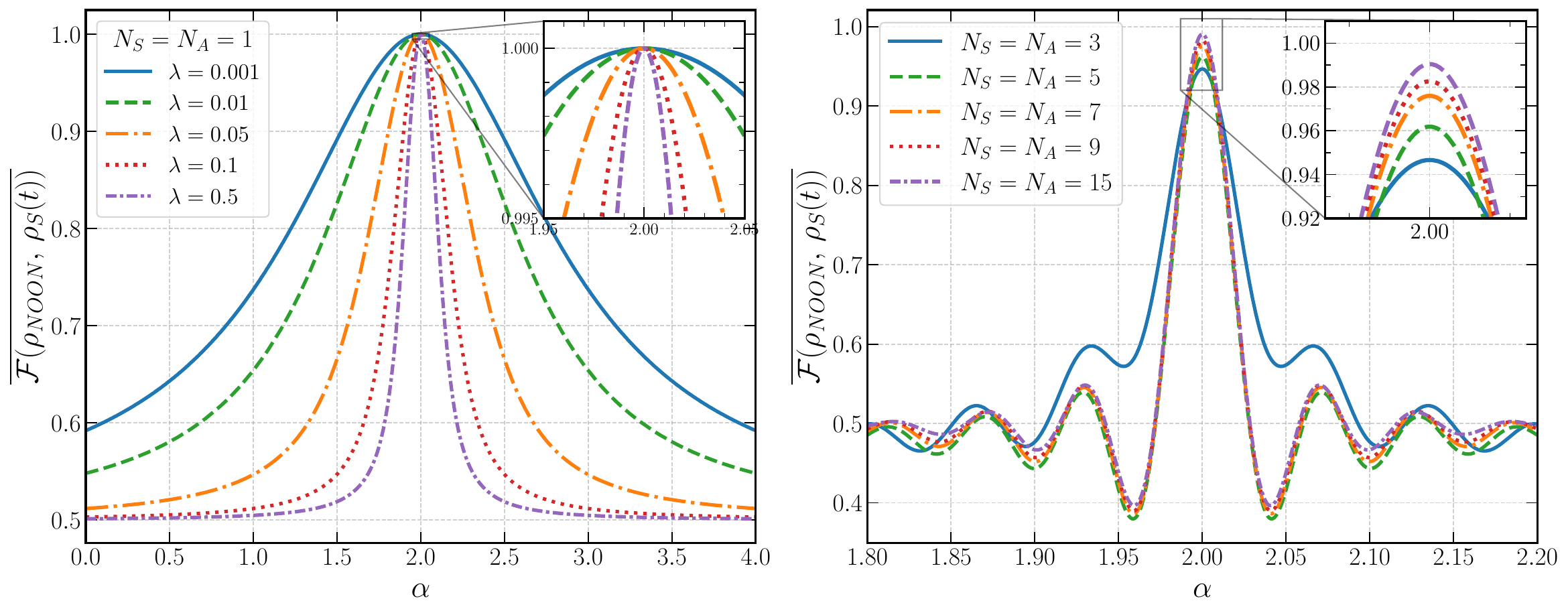}};
  \begin{scope}[x={(image.south east)},y={(image.north west)}]
    \node at (0.02,1) {\textbf{(a)}};
    \node at (0.52,1) {\textbf{(b)}};
  \end{scope}
\end{tikzpicture}

\caption{Time-averaged fidelity $\bar{\mathcal{F}}$ for preserving the NOON state  as a function of the interaction parameter $\alpha$. Each curve corresponds to a distinct  system–ancilla configuration with $N_S = N_A$, where the ancilla is prepared in the eigenstate $P_A^l$ of $J_A^z$ with $l = -\tfrac{1}{2}$. (a) For $N_S = N_A = 1$, the system reduces to a single qubit $S$, exhibiting a resonance peak centered at $\alpha_c = 2$. The resonance width decreases as the dephasing strength  $\lambda$ increases.
(b) For odd $N_S = N_A > 1$ and $\lambda=0.1$, $\bar{\mathcal{F}}$ displays a pronounced peak near $\alpha_c = 2$. The inset provides a magnified view, showing that the peak averaged fidelity approaches unity as the number of Bosons increases. Data for even $N_S = N_A$ follow the same trend and are omitted for clarity.} 
   \label{fig:Ns}
\end{figure*}

The protocol is robust against $\alpha\neq \alpha_c$. Indeed, given the fidelity
$\mathcal{F}(\alpha_,N)(t) = Tr\left[\rho_{SA}(t)\,\rho_{NOON}\otimes P^\ell_A\right]$, its time-average $\overline{\mathcal{F}(\alpha_,N)}$ signals how frequently, for a given $\alpha$, the time-evolving NOON state differs from itself. Figures~\ref{fig:Ns} display the behavior of $\overline{\mathcal{F}(\alpha_,N)} = \lim_{T\to\infty}\frac{1}{T}\int_{0}^{T}dt\,\mathcal{F}(\alpha_,N)(t)$ as function of $\alpha$ and for different values of $N$.
For any finite $N$, the residual discrepancy $
1-\overline{\mathcal{F}(\alpha_c,N)}
$ is due to the non-vanishing time–averaged contribution of the unitary oscillations generated by the effective Hamiltonian $H_S^{eff}$ in Eq.~\eqref{coupledGKSL8}. These residual oscillations scale as $1/N$ and thus vanish in the $N\to\infty$ limit.
The central width of the curves decreases with increasing strength $\lambda$ of the dephasing, as shown in Figure~\ref{fig:Ns}. Yet, the values of $\lambda$ consistent with the weak-coupling limit hypotheses underlying the derivation of the GKSL master equation from an actual system-environment interaction, the narrowness of the resonance is not dramatic. 

We note that the coupling constant $\alpha$ ought to be  optimally tuned depending on the initial state of the ancilla; therefore, for a not much tunable coupling $\alpha$ could be compensated by preparing the auxiliary system in an eigenstate of $J_A^z$ with eigenvalue $\ell$ closest to $-1/\alpha$. Analogously, the protocol is also robust against imperfect ancilla preparation, as shown in the Supplementary Section~\ref{sec:robustnessA_PRL}. Furthermore, by independently increasing the number of bosons $N_A$ in the ancillary sector, it is in principle possible to enhance the number of available dark channels for noise cancellation. Specifically, in the Jordan–Schwinger representation of the $\mathfrak{su}(2)$ algebra with $N_A$ bosons, the operator $J_A^z$ admits $2N_A - 1$ distinct eigenstates. Each of these can serve as a candidate ancillary dark state, thereby providing a larger set of interference channels that can be exploited to suppress noise more effectively.

We stress that, once the noise cancellation is achieved, the Hilbert space of the system is protected from the noise. The protocol thus appears to be scalable; indeed, it depends only on the ability to $1)$  prepare the initial state of the auxiliary system and $2)$ tune the $SA$-coupling strength $\alpha$.  As an illustration of this benefit in the case of the $S$ NOON state, Figure~\ref{fig:Ns} shows that increasing the dimension of $S$ and $A$ improves on its resilience against noise. We note, moreover, that the condition on state preparation of the ancilla can be relaxed. In fact, in Section~\ref{sec:robustnessA_PRL} of the SM it is shown that, if the ancilla is prepared in a mixed state with mean excitation $\bar \ell$ and width $\Delta \ell$, for white noise decoherence is cancelled when $\alpha=-1/\bar\ell$ and $\lambda\Delta^2\ell\ll 1$. 

In conclusion, the just presented proposal exploits dynamically generated internal noise interferences to protect from decoherence any state of a noisy system $S$. It differs from existing techniques as those based on the identification of decoherence-free subspaces or upon feed-back; rather, it resembles the mechanisms behind classical stochastic resonance by virtue of the mechanism of resonant response that leads to noise cancellation \cite{gammaitoni}. The method works for any spectral property of the noise, that is not only for white-noise and can provide a key strategy to perform high-precision metrology or sensing and to increase the quantum volume of a circuit when the noise is non-Markovian and typical quantum error-correction codes become inefficient.

\begin{acknowledgments}
 F.B. acknowledges financial support from PNRR MUR project PE0000023-NQSTI. We thank Andrea Trombettoni for insightful discussions regarding the critical aspects and possible experimental implementations of the proposed dephasing cancellation method. 
\end{acknowledgments}

\section*{Supplementary Material}
\section{GKSL Master equation from Stochastic processes} 
\label{SM:app1}

Let us consider the time-dependent Hamiltonian
\begin{equation} \label{Hamt}
H(t) \,= \, H_0 \,+ \,\bw(t)\cdot\bA\ ,\ \bw(t)\cdot\bA=\sum_{i=1}^nw_i(t)\, A_i\ , 
\end{equation}
where the operators $A_i=A^\dag_i$ commute among themselves, $H_0$ is a generic time-independent Hamiltonian, and the explicit time-dependence arises from correlated white-noises $w_i(t)$ such that 
\begin{equation} \label{white1}
\llangle w_i(\tau)\rrangle = 0\,, \ \llangle w_i(\tau)\, w_j(\tau') \rrangle \, = \, 
\Lambda_{ij} \,\delta(\tau-\tau')\ ,
\end{equation}
with $\Lambda=[\Lambda_{ij}]$ a positive $n\times n$ positive correlation matrix.

The stochastic potential gives rise to a Stratonovich stochastic Schr\"odinger equation 
\begin{equation} \label{msdli}
i\frac{d}{dt}\, \ket{\psi_{\bw}(t)}\, = \, \left[ H_{0} \;
+ \; \bw(t)\cdot\bA\,\right] .
\ket{\psi_{\bw}(t)}
\end{equation} 
For each realization of the noise, the formal solution of equation (\ref{msdli}) is:
\begin{equation} \label{cito}
\ket{\psi_{\bw}(t)} \; = \; T\, e^{\displaystyle -i H_{0}t
\; - \; i\,\int_{0}^{t}d\tau \,  \bw( \tau)\cdot\bA} \ket{\psi(0)}\ ,
\end{equation}
where $T$ is the usual time-ordering operator and $\ket{\psi(0)}$ the noise-independent initial condition. The state 
at time $t\geq0$ is obtained by
averaging over the noise:
\begin{equation} \label{ro}
\rho(t)\;=\,\llangle \ket{\psi_{\bw}(t)}\bra{\psi_{\bw}(t)} \rrangle \ .
\end{equation}
Considering $\rho(t+\epsilon)$ and expanding up to order one in $\epsilon$, we get~\cite{ghirardi1990relativistic,bassi2003dynamical}:
\begin{align} 
\label{mcl2}
\nonumber
\rho(t+\epsilon) & =  \NLLangle \,\Big[ 1 - iH_{0}\epsilon - i W_1(\epsilon)  -\frac{1}{2}\,W_2(\epsilon) \, +\, \mathcal{O}(\epsilon^2) \Big]
\\
\nonumber 
&  |\psi_{\bw}(t)\rangle\langle\psi_{\bw}(t)| \Big[ 1 + iH_{0}\epsilon + i W_1(\epsilon)  \\
&-\frac{1}{2}\,W_2(\epsilon) \, +\, \mathcal{O}(\epsilon^2) \Big] \,\NRRangle
\end{align}
where 
\begin{align} 
\label{WW}
W_1(\epsilon) & := 
\int_{t}^{t+\epsilon} \hskip-.5cm d\tau \,
\bw( \tau) \cdot\bA\\
\label{WW.1}
W_2(\epsilon)& :=  \sum_{i,j=1}^nA_iA_j\,\int_{t}^{t+\epsilon}\hskip-.5cm d\tau  \int_{t}^{t+\epsilon}
\hskip-.5cm d\tau' \, w_i(\tau)\, w_j( \tau') \ ,
\end{align}
and  only terms   possibly providing contributions of order $\mathcal{O}(\epsilon)$ have been included, while the factors $\frac{1}{2}$ come from applying the time-ordering to double time integrals.

We thus obtain, using (\ref{ro}):
\begin{widetext}
\begin{align} 
\nonumber
\rho(t+\epsilon) &=  \rho(t) \, -i\,\epsilon\ \big[ H_0, \rho(t)\big] -i\, \llangle\,\big[ \,W_1(\epsilon) ,\,\ket{\psi_{\bw}(t)}\bra{\psi_{\bw}(t)}\,\big] \,\rrangle -\frac{1}{2} \,\llangle\, \big\{ \,W_2(\epsilon),\,\ket{\psi_{\bw}(t)}\bra{\psi_{\bw}(t)}\big\} \,\rrangle\\
\nonumber
&+\epsilon \,\Big( \llangle\, W_1(\epsilon)\,\ket{\psi_{\bw}(t)}\bra{\psi_{\bw}(t)}\,\rrangle \,H_0 + H_0\,\llangle\, \,\ket{\psi_{\bw}(t)}\bra{\psi_{\bw}(t)}\,W_1(\epsilon)\rrangle \Big)\\
\nonumber
&-\frac{i\,\epsilon}{2} \,\Big( \,\llangle\, W_2(\epsilon)\,\ket{\psi_{\bw}(t)}\bra{\psi_{\bw}(t)}\,\rrangle \,H_0 -\,H_0\,\llangle\, \,\ket{\psi_{\bw}(t)}\bra{\psi_{\bw}(t)}\,W_2(\epsilon)\rrangle \Big)\\
\nonumber
&+\, \llangle\, \,W_1(\epsilon) \,\ket{\psi_{\bw}(t)}\bra{\psi_{\bw}(t)}\,W_1(\epsilon) \,\rrangle-\frac{i\,\epsilon}{2} \,\Big(\, \llangle\, W_2(\epsilon)\,\ket{\psi_{\bw}(t)}\bra{\psi_{\bw}(t)}\,W_1(\epsilon)\rrangle-\llangle\, W_1(\epsilon) \,\ket{\psi_{\bw}(t)}\bra{\psi_{\bw}(t)}\,W_2(\epsilon)\rrangle \Big)\\
\label{roteps2}
&+\, \frac{1}{4}\llangle\, \,W_2(\epsilon) \,\ket{\psi_{\bw}(t)}\bra{\psi_{\bw}(t)}\,W_2(\epsilon) \,\rrangle +\mathcal{O}(\epsilon^2)\ .
\end{align}
\end{widetext}
One then proceeds using 
the Furutsu-Novikov-Donsker relation: in the case of white-noise it reads
\begin{equation}
    \label{FurNovWhite}
    \llangle w_i(s) \,F[w(t)]\rrangle\,=\,\lambda\,\LLangle  \frac{\delta F[w(t)]}{\delta w_i(s)}\RRangle \ , 
\end{equation}
in terms of the functional derivative with respect to the noise.

Using (\ref{FurNovWhite}) and (\ref{WW}) we compute the term:
\begin{eqnarray}
    \label{W1psipsi}
    \nonumber
     &&\llangle \,W_1(\epsilon) \,\ket{\psi_{\bw}(t)}\bra{\psi_{\bw}(t)} \,\rrangle\\
     \nonumber
    && =\,\bA\cdot\int_{t}^{t+\epsilon} d\tau\, \llangle \,\bw(\tau) \,\ket{\psi_{\bw}(t)}\bra{\psi_{\bw}(t)} \,\rrangle\\
    &&=\,\lambda\,\bA\cdot\,\int_{t}^{t+\epsilon} d\tau\,
    \LLangle  \frac{\delta\ket{\psi_{\bw}(t)}\bra{\psi_{\bw}(t)}}{\delta\bw(\tau)}\RRangle  
\end{eqnarray}

The average in \eqref{W1psipsi} involves the functional derivative of a function of the noise at time $t$
with respect to the noise at time $\tau$, $w(\tau)$, within a time-integral over $[t,t+\epsilon]$. 
Terms like this vanish so that 
\begin{eqnarray}
    \label{W1comm}
    \nonumber
     &&\llangle\,\big[ \,W_1(\epsilon) ,\,\ket{\psi_{\bw}(t)}\bra{\psi_{\bw}(t)}\,\big] \,\rrangle\\
     \nonumber
     &&=\,\,\lambda\,\sum_{i=1}^n\left[A_i\,,\,\int_{t}^{t+\epsilon} d\tau\,
    \LLangle  \frac{\delta\ket{\psi_{\bw}(t)}\bra{\psi_{\bw}(t)}}{\delta w_i(\tau)}\RRangle  \right]=0\ .
\end{eqnarray}
Instead, using twice~\eqref{FurNovWhite}, double integrals, like
\begin{eqnarray}
    \label{W2psipsi}
    \nonumber
   &&\llangle \,W_2( \epsilon\ket{\psi_{\bw}(t)}\bra{\psi_{\bw}(t)}\rrangle\\
   \nonumber
   &&= \sum_{i,j=1}^nA_iA_j\int_{t}^{t+\epsilon} \hskip-.4cm d\tau  \int_{t}^{t+\epsilon}\hskip-.4cm d\tau'\llangle w_i(\tau)\,w_j(\tau') \ket{\psi_{\bw}(t)}\bra{\psi_{\bw}(t)}\rrangle,\\
\end{eqnarray}
lead to
\begin{equation}
\llangle \,W_2(\epsilon) \,\ket{\psi_{\bw}(t)}\bra{\psi_{\bw}(t)} \,\rrangle=\epsilon\,\sum_{i,j=1}^n\lambda_{ij}A_iA_j\,\rho(t)\ ,
    \label{W2psipsi.2}
\end{equation}
and, consequently, to contributions of order $\epsilon$ of the form
\begin{equation}
    \label{W2anticomm}
     \llangle\big\{W_2(\epsilon),\ket{\psi_{\bw}(t)}\bra{\psi_{\bw}(t)}\big\}\rrangle=\epsilon\sum_{i,=1}^n\lambda_{ij}
     \left\{A_iA_j,\rho(t) \right\}.     
\end{equation}
Analogously,
\begin{equation}
    \label{W1psipsiW1}
\llangle W_1(\epsilon)\ket{\psi_{\bw}(t)}\bra{\psi_{\bw}(t)}W_1(\epsilon)\rrangle
=\epsilon\sum_{i,j=1}^n\lambda_{ij}A_i\rho(t) A_j\ .
\end{equation}
Since all terms containing more than two integrations over $[t,t+\epsilon]$ vanish faster than $\epsilon$, the following master equation
ensues:
\begin{align} 
\nonumber
&\frac{d \rho(t) }{d t}=   -i\, \big[ H_0, \rho(t)\big]\\
\label{rodt}
&+\sum_{i,j=1}^n\lambda_{ij}\,\Big(A_i\,\rho(t) \,A_j\,-\,\frac{1}{2}\,\left\{A_jA_i\,,\,\rho(t) \right\}\Big)\ .
\end{align}

This expression represents the standard 
Gorini-Kossakowski-Sudarshan-Lindblad (GKSL) master equation governing the time evolution of $\rho(t)$, with the self-adjoint commuting operators $A_i$ serving as jump operators.

\section{Derivation of the master equation}
\label{SM:ME}

With $A_1=J^z_S\otimes \mathbb{I}_A$, $A_2=\mathbb{I}_S\otimes J^z_A$ and $A_3 = J^z_S\otimes J^z_A$ and the correlation matrix 
\begin{equation}
\label{corr-mat2}
  \Lambda=[\Lambda_{ij}]=\lambda\begin{pmatrix}
  1&1&\alpha\\
 1&1&\alpha\\
  \alpha&\alpha&\alpha^2
  \end{pmatrix}\ ,
  \end{equation}
using the previous section one finds a  GKSL master equation with generator 
$
\mathbb{L}= \sum_{i,j =1}^3\mathbb{L}_{ij}$ \,,
where the diagonal terms read
\begin{eqnarray}
\nonumber
    \mathbb{L}_{11}\rho_{SA}&=&\lambda\Bigg( J^z_S\otimes \mathbb{I}_A\,\rho_{SA}\,J^z_S\otimes \mathbb{I}_A\\
    \label{rodtapp1}
    &-&\,\frac{1}{2} \Big\{ (J^z_S)^2\otimes \mathbb{I}_A\,,\,\rho_{SA}\Big\}\Bigg)\ ,
\end{eqnarray}
\begin{eqnarray}
    \nonumber
    \mathbb{L}_{22}\rho_{SA}&=&\lambda \Bigg(\mathbb{I}_S\otimes J^z_A\,\rho_{SA}\,\mathbb{I}_S\otimes J^z_A\\
    \label{rodtapp2}
    &-&\,\frac{1}{2} \Big\{ \mathbb{I}_S\otimes (J^z_A)^2\,,\,\rho_{SA}\Big\}\Bigg)
    \ ,
\end{eqnarray}
\begin{eqnarray}
    \nonumber
    \mathbb{L}_{33}\rho_{SA}&=&\,\lambda\alpha^2\Bigg( J^z_S\otimes J^z_A\,\rho_{SA}\,J^z_S\otimes J^z_A\\
    \label{rodtapp3}
    &-&\,\frac{1}{2} \Big\{ (J^z_S)^2\otimes (J^z_A)^2\,,\,\rho_{SA}\Big\}\Bigg)\ ,
\end{eqnarray}
and the off-diagonal ones
\begin{eqnarray}
    \nonumber
    \mathbb{L}_{12}\rho_{SA}
    &=&\lambda\Bigg( J^z_S\otimes\mathbb{I}_A\,\rho_{SA}\,\mathbb{I}_S\otimes J^z_A\\
    \label{rodtapp4}
    &-&\,\frac{1}{2} \Big\{ J^z_S\otimes J^z_A\,,\,\rho_{SA}\Big\}\Bigg)\ ,
\end{eqnarray}
\begin{eqnarray}
    \nonumber
    \mathbb{L}_{21}\rho_{SA}
    &=&\lambda\Bigg( \mathbb{I}_S\otimes J^z_A\,\rho_{SA}\,J^z_S\otimes\mathbb{I}_A\\
    \label{rodtapp4b}
    &-&\,\frac{1}{2} \Big\{ J^z_S\otimes J^z_A\,,\,\rho_{SA}\Big\}\Bigg)\ ,
\end{eqnarray}
\begin{eqnarray}
    \nonumber
    \mathbb{L}_{13}\rho_{SA}
    &=&\lambda\alpha\Bigg( J^z_S\otimes\mathbb{I}_A\,\rho_{SA}\,J^z_S\otimes J^z_A\\
    \label{rodtapp4c}
    &-&\,\frac{1}{2} \Big\{ (J^z_S)^2\otimes J^z_A\,,\,\rho_{SA}\Big\}\Bigg)\ ,
\end{eqnarray}
\begin{eqnarray}
    \nonumber
    \mathbb{L}_{31}\rho_{SA}
    &=&\lambda\alpha\Bigg( J^z_S\otimes J^z_A\,\rho_{SA}\,J^z_S\otimes\mathbb{I}_A\\
    \label{rodtapp5}
    &-&\,\frac{1}{2} \Big\{ (J^z_S)^2\otimes J^z_A\,,\,\rho_{SA}\Big\}\Bigg)\ ,
\end{eqnarray}
and
\begin{eqnarray}
    \nonumber
    \mathbb{L}_{23}\rho_{SA}
    &=&\lambda\alpha\Bigg( \mathbb{I}_S\otimes J^z_A\,\rho_{SA}\,J^z_S\otimes J^z_A\\
    \label{rodtapp6}
    &-&\,\frac{1}{2} \Big\{J^z_S\otimes (J^z_A)^2\,,\,\rho_{SA}\Big\}\Bigg)\ ,
\end{eqnarray}
\begin{eqnarray}
    \nonumber
    \mathbb{L}_{32}\rho_{SA}
    &=&\lambda\alpha\Bigg( J^z_S\otimes J^z_A\,\rho_{SA}\mathbb{I}_S\otimes J^z_A\\
    \label{rodtapp7}
    &-&\,\frac{1}{2} \Big\{J^z_S\otimes (J^z_A)^2\,,\,\rho_{SA}\Big\}\Bigg)\ .
\end{eqnarray}
Then, in Eq. (8) of the main text, $\mathbb{M}[\hat O_1]=\mathbb{L}_{11}+\mathbb{L}_{13}+\mathbb{L}_{31}+ \mathbb{L}_{33}$, 
$\mathbb{M}[\hat O_2]=\mathbb{L}_{22}+\mathbb{L}_{23}+\mathbb{L}_{32}$, and $\mathbb{L}_{12}+\mathbb{L}_{21}=\mathbb{M}[\hat O_{+}]-\mathbb{M}[\hat O_{-}]$.

\section{Physical implementation}
\label{sec:physimpl}

As emphasized in the main text,  the proposed methodology for mitigating decoherence demands two conditions to be satisfied: 1) that, when target system $S$ and ancilla $A$ are not coupled, the dephasing white noises 
affecting them extend to a correlated white noise; 2) that, when switched on,  also the coupling term $J_S^z\otimes J_A^z$ becomes noisy. We now sketch a possible concrete implementation of the proposed protocol, leaving the discussion of the  necessary physical details for a later dedicated paper.

Consider an extended Bose-Hubbard lattice model with long-range interactions~\cite{landig2016quantum, habibian2013bose, ritsch2013cold, klinder2015observation}: 
\begin{eqnarray}
\label{long-range_H}
\nonumber
    H &=& -t \sum_{\langle e,o \rangle} \left( b^\dagger_e b_o + \text{h.c.} \right)
+ \frac{U_s}{2} \sum_{i \in e,o} n_i (n_i - 1)\\
&&- \frac{U_1}{K} \left( \sum_{e} n_e - \sum_{o} n_o \right)^2 - \sum_{i \in e,o} \mu_i n_i.
\end{eqnarray}
Here, $b^\dagger_i$ and $b_i$ are bosonic creation and annihilation operators at site $i$, with $n_i = b^\dagger_i b_i$ the corresponding number operators. Even and odd lattice sites are denoted by $e$ and $o$, respectively, $K$ being their total number. The parameters $t$, $U_s$, and $U_1$ represent the nearest-neighbor tunneling amplitude, on-site interaction strength, and infinite-range interaction strength, respectively. The chemical potential at each site is $\mu_i = \mu - \varepsilon_i$, where $\mu$ is the global chemical potential and $\varepsilon_i$ an external potential offset.

\begin{figure}[t!]
\centering
\includegraphics[width=0.55\linewidth]{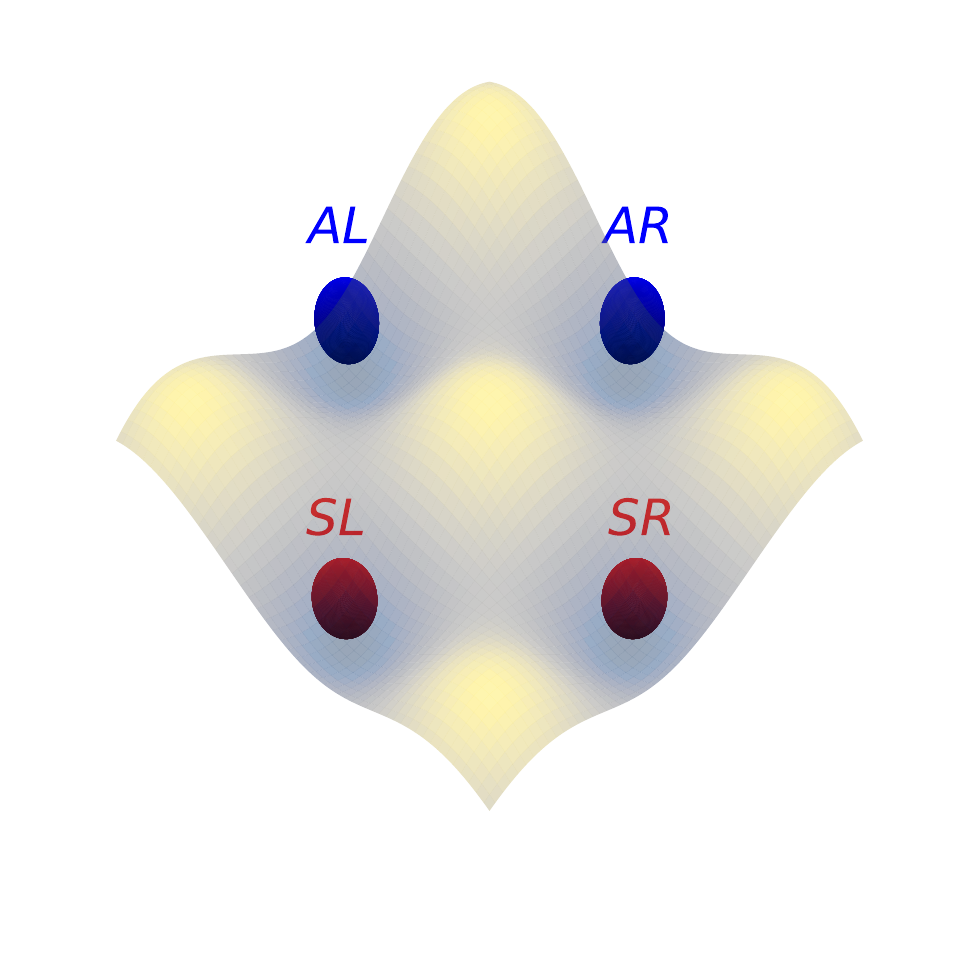}
%\put(-240,150){(a)}  % Adjust coordinates to position label 
\caption{Schematic representation of the four-sites plaquette. }
   \label{fig:plaquette}
\end{figure}

We now specialize the above general framework to consist of a four-site plaquette, comprising two double-wells, each trapping $N$ Bosons, as depicted in Fig~\ref{fig:plaquette}: there are two pairs $\langle e,o\rangle$, one for the system ($S$) and one for the ancilla ($A$), each with a left ($e=L$) and right ($o=R$) site.

Further, we also distinguish between the hopping probabilities for the $S$, respectively $A$ trap. 
Indeed, the protocol presented in the previous section requires the suppression of the tunneling probability ($t_A=0$) in the ancillary double-well. 

Then, the Bose-Hubbard like Hamiltonian becomes 
\begin{eqnarray}
\nonumber
H_{SA}&=&
-\,t_S \, b^\dagger_{\text{SR}} b_{\text{SL}} -\,t_A\,
 b^\dagger_{\text{AR}} b_{\text{AL}} + \text{h.c.}   \\
 \nonumber
&+&\frac{U_S}{2} \Big[ n_{\text{SR}} (n_{\text{SR}} - 1) 
+ n_{\text{AR}} (n_{\text{AR}} - 1) \\
&+& n_{\text{SL}} (n_{\text{SL}} - 1) 
+ n_{\text{AL}} (n_{\text{AL}} - 1) \Big]\nonumber \\
&+& \frac{U_1}{4} \big( n_{\text{SR}} + n_{\text{AR}} 
- n_{\text{SL}} - n_{\text{AL}} \big)^2 \nonumber \\
\label{ourH}
&&\hskip -1.5cm-\big( \, \mu_{\text{SR}}\, n_{\text{SR}} + \mu_{\text{AR}} \,n_{\text{AR}}+ \mu_{\text{SL}} \,n_{\text{SL}} + \mu_{\text{AL}}\, n_{\text{AL}} \big) \ .
\end{eqnarray}
By using the Jordan-Schwinger representation of the $\mathfrak{su}(2)$ algebra with $N_A$ bosons, one introduces angular-momentum operators as follows:
\begin{align*}
&
b^\dag_{iL}b_{iL}:=n_{\text{iL}}=\frac{N}{2}-J^z_i\ ,\ b^\dag_{iR}b_{iR}:=n_{\text{iR}}=\frac{N}{2}+J^z_i\ ,\\
&
\frac{b_{iL}b^\dag_{iR}+b^\dag_{iL}b_{iR}}{2}=J^x_i\ ,\ 
\frac{b_{iL}b^\dag_{iR}-b^\dag_{iL}b_{iR}}{2i}=J^y_i\ .
\end{align*}
Notice that the eigenstate $\ket{\ell}$ of 
$\displaystyle J_i^z=\frac{n_{\text{iR}}-n_{\text{iL}}}{2}$ 
with eigenvalue $-\frac{N}{2}\leq\ell\leq \frac{N}{2}$ corresponds to a state with $\frac{N}{2}-\ell$ Bosons in the left well of the $i$-th trap and $\frac{N}{2}+\ell$ Bosons in the right one.
To express the Hamiltonian in a more convenient form, we introduce the hopping strengths $\gamma_i = 2\,t_i$ for $i = S, A$, along with the energy asymmetries $\Delta_S = \mu_{\text{SR}} - \mu_{\text{SL}}$ and $\Delta_A = \mu_{\text{AR}} - \mu_{\text{AL}}$. We further set the onsite Coulomb interaction to $\eta = \eta_S = \eta_A = U_s$ and define the long-range interaction as $\alpha = -2\,U_1$. With these definitions, the Hamiltonian~\eqref{ourH} can be recast, up to a multiple of the identity, into the form
\begin{eqnarray}
\nonumber
H_{SA} &=& 
 -\gamma_S\,J_S^x \otimes \mathbb{I}_A  -\gamma_A\,\mathbb{I}_S \otimes J_A^x\\
\nonumber 
&-&  \Delta_S \,J_S^z\otimes \mathbb{I}_A - \Delta_A \,\mathbb{I}_S \otimes J_A^z+ \alpha J_S^z \otimes J_A^z 
 \\
 \label{ourH.2}
 &&+ \left( \eta + \frac{\alpha}{2} \right) \Big( (J_S^z)^2\otimes \mathbb{I}_A  + \mathbb{I}_S\otimes (J_A^z)^2 \Big) \ .
\end{eqnarray}
If the chemical potentials are stochastically perturbed by white-noises as follows,
\begin{equation}
\label{noises}
\mu_{iR} \rightarrow \mu_{iR} + w_i(t)/2\ ,\quad \mu_{iL} \rightarrow \mu_{iL} - w_i(t)/2\ ,
\end{equation}
the asymmetry parameters transform as:
\begin{equation}
\label{delta_pert}
\Delta_{i}\rightarrow \Delta_{i} + w_i(t) \, \, \qquad i=S,A\ .
\end{equation}
To ensure effective noise cancellation, we super-impose a correlated stochastic modulation upon the interaction coupling $U_1$, such that:
\begin{equation}
\label{alfa_pert}
\alpha\rightarrow \alpha\big(1 + w_{SA}(t)\big)  .
\end{equation}
As a result, the perturbed Hamiltonian takes the form:
\begin{eqnarray}
\nonumber
\hspace{-5mm}H_{SA}(t) &=& 
-\gamma_S\,J_S^x \otimes \mathbb{I}_A  -\,\gamma_A\, \mathbb{I}_S \otimes J_A^x\\
\nonumber 
&&-  \Delta_S \,J_S^z\otimes \mathbb{I}_A- \Delta_A \,\mathbb{I}_S \otimes J_A^z + \alpha J_S^z \otimes J_A^z\\
 \nonumber
 &&+ \left( \eta + \frac{\alpha}{2} \right) \Big( (J_S^z)^2\otimes \mathbb{I}_A  + \mathbb{I}_S\otimes (J_A^z)^2 \Big) \\
 \nonumber
 && + w_S(t)\,J_S^z\otimes \mathbb{I}_A +w_A(t)\mathbb{I}_S \otimes J_A^z\\
\nonumber 
&& + \alpha\, w_{SA}(t)\Bigg( J_S^z \otimes J_A^z\\
\label{ourH.4}
&&+\frac{1}{2} \Big((J_S^z)^2\otimes \mathbb{I}_A  + \mathbb{I}_S\otimes (J_A^z)^2 \Big)\Bigg)\ .
\end{eqnarray}

The key tool in the above protocol is the driving of the chemical potentials and of the $U_1$ coupling by 
noises that are correlated according to~\eqref{corr-mat}.

The parameters $U_1$ and $\mu_i$ could originate from optical fields, and the noise associated with them from intensity or frequency fluctuations. Specifically, $U_1$ describes processes where a laser photon is coherently scattered into the cavity mode. Let $\Omega$ be the laser Rabi frequency, $g$ be the vacuum Rabi frequency, and $\Delta$ the de-tuning from atomic resonance, then $U_1\,=\,g\,\Omega/\Delta$. Correspondingly, the atoms experience a frequency shift $\mu_i=\Omega^2/\Delta$, as well as a shift $g^2/\Delta$ per cavity photon. Noise in the laser frequency is a stochastic fluctuation of $\Delta$, which equally affects both $U_1$ and $\mu_i$. Noise in the laser intensity gives rise to fluctuations of $\Omega$, thus giving rise to correlated fluctuations of the two parameters. 

Also in ZZ-interacting capacitively coupled superconducting transmon qubits~\cite{long2021universal}, the transmon frequencies and the coupling may be perturbed following a similar protocol, exploiting the ZZ-interaction instead of trying to suppress it.

\subsection{A case-study}
\label{App:case-study}

Once the tunneling in the ancilla double-well is suppressed and $\gamma_A=0$, compared with the Hamiltonian
\begin{align}
   \nonumber
    H_{SA}(t)&=\,H_0\, +\, w_S(t)\,J^z_S\otimes \mathbb{I}_A\,+\,w_A(t)\, \mathbb{I}_S\otimes J^z_A\\ 
     \label{HSA2}
     &+\,w_{SA}(t)\,J^z_S\otimes J^z_A\ ,
\end{align}
(see Eq. (4) in the main text), the Hamiltonian in~\eqref{ourH.4} shows an extra term 
\begin{equation}
\label{extra1}
K_{SA}:=\frac{(J_S^z)^2\otimes \mathbb{I}_A  + \mathbb{I}_S\otimes (J_A^z)^2}{2}\ .
\end{equation}    
This latter operator contributes to the GKSL generator $\mathbb{L}$ in Eq. (6) of the main text with an additional term
\begin{align}
\nonumber
\mathbb{M}[\hat{O}_3]\rho_{SA}&=\lambda\alpha\Big(J^z_S\otimes \mathbb{I}_A\,\rho_{SA}\,K_{SA}\\
\label{extra2a}
&-\frac{1}{2}\Big\{K_{SA}J^z_S\otimes\mathbb{I}_A\,,\,\rho_{SA}\Big\}\Big) +h.c.\\
\nonumber&+\lambda\alpha\Big((\mathbb{I}_S\otimes J^z_A\,\rho_{SA}\,K_{SA}\\
&-\frac{1}{2}\Big\{K_{SA}J^z_S\otimes\mathbb{I}_A\,,\,\rho_{SA}\Big\}\Big) +h.c.
\label{extra2b}\\
\label{extra2c}
&\hskip-1cm
+\lambda\alpha^2\Big(K_{SA}\,\rho_{SA}\,K_{SA}-\frac{1}{2}\Big\{(K_{SA})^2\,,\,\rho_{SA}\Big\}\Big)\ .
\end{align}
If the state of the ancilla is chosen as $P^\ell_A$ with $\alpha=-\frac{1}{\ell}$, then $\mathbb{M}[\hat{O}_1]\rho_S\otimes P^\ell_A
=\mathbb{M}[\hat{O}_2]\rho_S\otimes P^\ell_A=0$. However, since $\mathbb{M}[\hat{O}_3]\rho_S\otimes P^\ell_A\neq 0$, residual dissipation still
affects the dynamics of $\rho_S$.
None the less, there are cases where no dissipation is induced by the presence of the contribution $K_{SA}$ in~\eqref{extra1}: as discussed in the main text, one of them is provided by the NOON-state; another example is when $N_S=N_A=1$, where the $J_S^i$ and $J_A^i$ operators reduce to the usual Pauli matrices and thus terms proportional to  $(J_S^z)^2$ and $(J_A^z)^2$ become negligible scalars. Notably in this latter case the cancellation scheme works for any initial qubit $\rho_S$.

\subsection{Laser-Induced White Noise and the Correlation Matrix
  \texorpdfstring{$\Lambda$}{Lambda}}
\label{laser}

Let the system be driven by a single classical laser field of (slowly varying) complex amplitude
\begin{equation}
\Omega(t) = \Omega_0 [1 + \xi_\Omega(t)] e^{i\phi_0},
\end{equation}
where $\xi_\Omega(t)$ represents intensity fluctuations, and $\phi_0$ is a constant laser phase.

In the far-detuned regime $|\Delta| \gg g,\,\Omega_0,\,\kappa$, where $|\Delta|$ is the laser--atom detuning, $g$ is the atom--cavity coupling strength, $\Omega_0$ is the Rabi frequency, and $\kappa$ is the cavity decay rate, the effective parameters read
\begin{eqnarray}
U_1(t) &=& \frac{g\,\Omega(t)}{\Delta}, \\
\mu_i(t) &=& \frac{|\Omega(t)|^2}{\Delta} + (-1)^{p(i)}\frac{g^2}{\Delta}, \label{mu_i}
\end{eqnarray}
with $p(i)=0,1$ for even/odd sites.  The first term in~\eqref{mu_i} is the usual laser-induced AC Stark shift of the atomic ground state; the second term alternates in sign on different lattice sites because of the standing-wave cavity mode.

Linearising to first order in the small real fluctuations $|\xi_{\Omega}|\ll1$, we obtain
\begin{eqnarray}
\delta\mu_S(t) = \delta\mu_A(t) &=& \frac{2\,\Omega_0^2}{\Delta}\,\xi_{\Omega}(t), \\
\delta U_1(t) &=& \frac{g}{\Delta}\,\xi_{\Omega}(t)\,e^{i\phi_0}.
\end{eqnarray}
Tuning $\phi_0$ such that only the real part of $\delta U_1(t)$ remains, the three classical stochastic couplings entering the Hamiltonian become
\begin{equation}
\mathbf{w}(t) = \sqrt{\lambda}\,
\begin{pmatrix}
 c_S \\
 c_S \\
 c_{SA}
\end{pmatrix}
\eta(t),
\end{equation}
where $\xi_{\Omega}(t)=\sqrt{\lambda}\,\eta(t)$, $\eta(t)$ is a real stochastic process with $\llangle\eta(t)\eta(t')\rrangle=f(t,t')$ and
\begin{equation}
c_S = \frac{2\,\Omega_0^2}{\Delta}, \quad c_{SA} = \frac{g}{\Delta}\cos\phi_0.
\end{equation}

Introducing the dimensionless ratio
\begin{equation}
\alpha = \frac{c_{SA}}{c_S} = \frac{g}{2\,\Omega_0^2}\cos\phi_0,
\end{equation}
and absorbing the common prefactor $\lambda c_S^2$ into $\lambda$,  the  second moments read
\begin{eqnarray}
\nonumber
\llangle w_i(t)w_j(t')\rrangle &=& \Lambda_{ij}\,f(t,t') \quad \hbox{where}\\
\Lambda=[\Lambda_{ij}]&=&\begin{pmatrix}
1 & 1 & \alpha \\
1 & 1 & \alpha \\
\alpha & \alpha & \alpha^2
\end{pmatrix}
\label{eq:corr}
\end{eqnarray}
is a positive semi-definite matrix for all $\lambda>0$ and $\alpha\in\mathbb{R}$, ensuring complete positivity of the resulting GKSL generator.
Moreover, the positive function $f(t,t')$ allows to consider not only white-noises, when $f(t,t')=\delta(t-t')$, but also colored and even $1/f$ noises.

\paragraph{Experimental Implementation}
The correlator above arises naturally by feeding the acousto-optic modulator (AOM) controlling the laser power with broadband electronic noise.  The overall RF power sets $\lambda\propto P_{\mathrm{RF}}$, while $\alpha$ is tuned via the mean Rabi frequency $\Omega_0$, the vacuum Rabi splitting $g$, or the phase $\phi_0$.  Commercial modulators with bandwidths $\gtrsim 100\,\mathrm{MHz}$ guarantee the white-noise assumption holds on the relevant dynamical timescales, and the noise spectra can be calibrated in situ with a fast photodiode and vector spectrum analyzer.

\section{Robustness to imperfect ancilla preparation}
\label{sec:robustnessA_PRL}

In a realistic implementation of the protocol, the ancilla $A$ cannot be perfectly
prepared in the exact eigenstate $\ket{\ell}_A$ of $J_A^z$ but only in a
narrowly–localised mixture around it.
We model the ensuing uncertainty by the diagonal state  
\begin{eqnarray}
\rho_{A}&=&\sum_{n=0}^{\infty}p_{n}\,\ket{n}\bra{n}_{A},
\nonumber\\
p_{n}&=&
\frac{\exp\!\bigl[-(n-\mu)^{2}/(2\sigma^{2})\bigr]}
{\displaystyle\sum_{m=0}^{\infty}\exp\!\bigl[-(m-\mu)^{2}/(2\sigma^{2})\bigr]},
\label{eq:rhoA}
\end{eqnarray}
with mean $\mu=\ell+\delta$ ($|\delta|\ll\ell$) and variance $\sigma^{2}\ll\ell^{2}$.  
Expectation values therefore read  
\begin{eqnarray}
  \langle J_A^z\rangle &:=& Tr[\rho_A J_A^z] = \mu,
  \nonumber\\
  \langle (J_A^z)^2\rangle &:=& Tr[\rho_A (J_A^z)^2] = \mu^2 + \sigma^2.
\end{eqnarray}

At $t=0$ the composite state is $\rho_{SA}(0)=\rho_{S}(0)\otimes\rho_{A}$.  
Choosing the dark–state condition $\alpha=-1/\ell$ the two–body jump operators  
\begin{eqnarray}
O_{1}&=&\sqrt{\lambda}\,J^{z}_{S}\otimes(\mathbb{I}_A+\alpha J^{z}_{A}),
\nonumber\\
O_{2}&=&\sqrt{\lambda}\,(\mathbb{I}_S+\alpha J^{z}_{S})\otimes J^{z}_{A},
\end{eqnarray}
yield, after tracing out $A$,  
\begin{eqnarray}
Tr_{A}\!\bigl[\mathbb{M}[O_{1}](\rho_{S}\!\otimes\!\rho_{A})\bigr]
&=&
\lambda\!\Bigl(1-\frac{\mu}{\ell}\Bigr)^{2}\,\mathbb{M}[J^{z}_{S}]\,\rho_{S},
\label{eq:R1}\\
Tr_{A}\!\bigl[\mathbb{M}[O_{2}](\rho_{S}\!\otimes\!\rho_{A})\bigr]
&=&
\lambda\,\frac{\mu^{2}+\sigma^{2}}{\ell^{2}}\,\mathbb{M}[J^{z}_{S}]\,\rho_{S}.
\label{eq:R2}
\end{eqnarray}
In the ideal limit $\mu=\ell$ these two–body terms alone would sum to
$\lambda\,\mathbb{M}[J_S^z]\rho_S$, but our correlated‐noise protocol
exactly cancels that $\lambda$ contribution, leaving only the
ancilla–preparation correction proportional to the small parameters
$\delta,\sigma$.
The single–body combinations  
$O_{\pm}=\sqrt{\lambda/2}\,(J^{z}_{S}\otimes\mathbb{I}_A\pm\mathbb{I}_S\otimes J^{z}_{A})$  
cancel exactly and introduce no additional dephasing.
Retaining the coherent contribution $-i\,[H^{\mathrm{eff}}_{S},\rho_{S}]$ one obtains the effective master equation
\begin{equation}
\partial_{t}\rho_{S}
=
-i[H^{\mathrm{eff}}_{S},\rho_{S}]
+
\Gamma_{\mathrm{res}}^{(A)}\,\mathbb{M}[J^{z}_{S}]\,\rho_{S},
\label{eq:master}
\end{equation}
where the residual dephasing rate is
\begin{equation}
\Gamma_{\mathrm{res}}^{(A)}
=\lambda\Bigl[\Bigl(1-\tfrac{\mu}{\ell}\Bigr)^{2}
+\tfrac{\sigma^2}{\ell^2}\Bigr]
=\lambda\,\frac{\delta^2+\sigma^2}{\ell^2}.
\label{eq:Gamma}
\end{equation}
where $\;|\delta|,\sigma\ll\ell\;$.

We define the coherence time 
$
\tau^A:=
1/(\varepsilon\,\Gamma_{\mathrm{res}}^{(A)})$.

\paragraph{Optical–tweezer benchmarks.}

State–of–the–art tweezer arrays report single–atom preparation fidelities  $\gtrsim99.97\%$ ($\varepsilon\lesssim3\times10^{-4}$) with survival probabilities beyond $99.8\%$ over thousands of imaging cycles \cite{tao2024high}. 
For $\ell=\tfrac12$, $\lambda=0.1\,\mathrm s^{-1}$, ancilla infidelity
$\varepsilon=3\times10^{-4}$ (so $\sigma^2\approx1.5\times10^{-4}$) and
systematic offset $\delta=0.01$, Equation~(\ref{eq:Gamma}) gives
\begin{equation}
\Gamma_{\mathrm{res}}^{(A)}
= 8\times10^{-5}~\mathrm s^{-1},
\end{equation}
implying
\begin{equation}
\tau\simeq 1.25\times10^{4}~\mathrm s\approx3.5~\mathrm{h}.
\end{equation}
Consequently, even allowing small calibration errors, the protocol suppresses
dephasing down to the $\mathcal O(\lambda\delta^2)$ level, achieving
multi‐hour coherence in current optical‐tweezer setups.

\section{Decoherence cancellation: Generic noises}
\label{SM:Dech_canc}

For generic noises, in the weak‐coupling regime,  at lowest non‐trivial order in the interaction picture, the generator of the dynamics is governed by the averaged double commutator $ \llangle [H_{SA}(t),\,[H_{SA}(t'),\,\rho_{SA}]]\rrangle$. 

Because the ancilla projector satisfies $J_A^z P_A^\ell=\ell\,P_A^\ell$, every noise source attached only to the ancilla cancels
\begin{eqnarray}
\nonumber
&&\left\langle \left[ H_{SA}(t), \left[ H_{SA}(t'), \rho_{SA} \right] \right] \right\rangle \\
&&= [H_0, [H_0, \rho_{SA}]] + C_\ell(t,t') [J_S^z, [J_S^z, \rho_S]] \otimes P_A^\ell 
\quad 
\label{eq:final_noise_average}
\end{eqnarray}
with second order correlator of the form
\begin{equation}
C_\ell(t,t')=\llangle\bigl[w_S(t)+\,\ell\,w_{SA}(t)\bigr]\,
\bigl[w_S(t')+\ell\,w_{SA}(t')\bigr]\rrangle\ .
\end{equation}
Assuming
\begin{eqnarray}
\llangle w_S(t) \,w_S(t') \rrangle &=&\lambda_S \,f(t,t')\\
\llangle w_{S}(t)\, w_{SA}(t') \rrangle &=&\lambda_{S,SA} \,f(t,t')
\\
\llangle w_{SA}(t)\, w_{SA}(t') \rrangle &=&\lambda_{SA,SA} \, f(t,t') \ ,
\end{eqnarray}
the second‐order correlator $C_\ell(t,t')$ vanishes if 
\begin{equation}
\lambda_{S} \;+\; 2\,\ell\,\lambda_{S,SA}\;+\;\ell^2\,\lambda_{SA,SA}
\;=\;0 \ .
\end{equation}
 Moreover, if as shown in Section~\ref{laser}, 
 one makes $\lambda_S =\lambda$, $\lambda_{S,SA}= \lambda\, \alpha$, and $\lambda_{SA,SA}=\lambda \,\alpha^2$, one retrieves the condition $\alpha_C =-1/\ell$ as in the main text.
Then, in the weak-coupling, there is no dephasing whatever the function $f(t,t')$ is.
Therefore, it holds for white, colored and even $1/f$ noises .

\bibliographystyle{apsrev4-1}
\bibliography{bibliography.bib}

\end{document}